\documentclass[12pt]{article}
\usepackage{amsfonts}
\usepackage{amsmath}
\usepackage{amssymb}
\usepackage{graphicx}
\usepackage{color}
\usepackage[all, knot]{xy}
\usepackage{tikz}

\usepackage[utf8]{inputenc}
\usepackage{subcaption}
\usepackage{epstopdf}
\usepackage{amsthm}
\usepackage{enumitem}
\usepackage{mathrsfs}

\usepackage[margin=3cm]{geometry}

\def \be {\begin{equation}}
\def \ee {\end{equation}}
\def \bea {\begin{eqnarray}}
\def \eea {\end{eqnarray}}
\def \nn {\nonumber}

\def \rr {\raise.35ex\hbox{\small $\prime$}\kern-.17em{\mbox{\large $\imath$}}}

\def \dels {\partial\kern-.6em /\kern.1em}
\def \As {{A\kern-.5em / \kern.5em}}
\def \Ds {D\kern-.7em / \kern.5em}

\def \ks {k\kern-.5em /}
\def \ls {l\kern-.5em /}

\newcommand{\ci}[1]{}
\newcommand{\ba}{\begin{eqnarray}}
\newcommand{\ea}{\end{eqnarray}}
\newcommand{\bal}{\begin{align}}
\newcommand{\eal}{\end{align}}
\newcommand{\bay}[1]{\left(\begin{array}{#1}}
\newcommand{\eay}{\end{array}\right)}

\newcommand{\hide}[1]{}

\DeclareMathOperator{\Tr}{Tr}

\newlist{axioms}{enumerate}{2}
\setlist[axioms,1]{label=\textbf{A\arabic{axiomsi}.}, ref=A\arabic{axiomsi}}
\setlist[axioms,2]{label=\textbf{A\arabic{axiomsi}\rlap{\myEnumCounter{axiomsii}}.},%
                   ref=A\arabic{axiomsi}\myEnumCounter{axiomsii},%
                   align=parleft,%
                   leftmargin=0em,%
                   itemsep=1.4ex,%
                   before={\stepcounter{axiomsi}}}
                   
  \usetikzlibrary{decorations.markings}

\begin{document}

\begin{titlepage}
\begin{center}

\textbf{\LARGE
Randomness and Chaos in Qubit Models
\vskip.3cm
}
\vskip .5in
{\large

Pak Hang Chris Lau$^{a,b}$ \footnote{e-mail address: phcl2@mit.edu}, Chen-Te Ma$^{c,d,e}$ \footnote{e-mail address: yefgst@gmail.com},\\ 
Jeff Murugan$^{d, f}$ \footnote{e-mail address: jeff.murugan@uct.ac.za}, and Masaki Tezuka$^g$ \footnote{e-mail address: tezuka@scphys.kyoto-u.ac.jp} 
\\
\vskip 1mm
}
{\sl
$^a$
National Center for Theoretical Sciences, National Tsing-Hua University,\\
Hsinchu 30013, Taiwan, R.O.C..
\\
$^b$
Center for Theoretical Physics, Massachusetts Institute of Technology,
Cambridge, MA02139, USA.
\\
$^c$
School of Physics and Telecommunication Engineering,\\
South China Normal University, Guangzhou 510006, China.
\\
$^d$
The Laboratory for Quantum Gravity and Strings,\\ 
Department of Mathematics and Applied Mathematics,
University Of Cape Town,
Private Bag, Rondebosch, 7701, South Africa.
\\
$^e$
Department of Physics and Center for Theoretical Sciences, \\
National Taiwan University, Taipei 10617, Taiwan, R.O.C..
\\
$^f$
Kavli Institute for Theoretical Physics University of California, Santa Barbara,\\
 CA 93106, USA.
\nn\\
$^g$
Department of Physics, Kyoto University, Kyoto 606-8502, Japan.
}\\
\vskip 1mm
\vspace{40pt}
\end{center}
\newpage
\begin{abstract}
We introduce randomness into a class of integrable models and study the spectral form factor as a diagnostic to distinguish between randomness and chaos. Spectral form factors exhibit a characteristic dip-ramp-plateau behavior in the $N>2$ SYK$_2$ model at high temperatures that is absent in the $N=2$ SYK$_2$ model. Our results suggest that this dip-ramp-plateau behavior implies the existence of random eigenvectors in a quantum many-body system. To further support this observation,  we examine the Gaussian random transverse Ising model and obtain consistent results without suffering from small $N$ issues. Finally, we demonstrate numerically that expectation values of observables computed in a random quantum state at late times are equivalent to the expectation values computed in the thermal ensemble in a Gaussian random one-qubit model.
\end{abstract}
\end{titlepage}

\section{Introduction}
\label{sec:1}

By now it is universally accepted that quantum mechanics is the fundamental theory that describes our world at low energies and atomic scales. On the other hand, the macroscopic world of everyday scales is very well described by classical Newtonian physics. Understanding the details of the transition from quantum to classical phenomena is of crucial importance, not only in various physical settings, like the early universe or black holes \cite{Wald:1984rg}, but also in various biological processes, such as photosynthesis \cite{Bryant:2006}. 
 One such phenomenon is {\it chaos}. Even though both classical and quantum chaotic systems have been the subject of much scrutiny over the past fifty or so years, the transition between an inherently quantum chaotic system and a corresponding classically chaotic system remains an important open problem. Largely, this is because tractable quantum systems that exhibit chaos are extraordinarily rare.\\

\noindent 
It is not surprising then, that the Sachdev-Ye-Kitaev (SYK) model of $N\gg1$ Majorana fermions with all-to-all random interactions have dominated both high energy and condensed matter physics for the past
few years since its discovery  \cite{Kitaev:2015,Sachdev:1992fk,Maldacena:2016hyu}. 
As a (0+1)-dimensional quantum mechanical system, it possesses three remarkable properties that distinguish it from 
the plethora of many-body systems \cite{Ma:2018efs}: {\it (i)} it exhibits an {\bf emergent conformal symmetry} at low energies, 
{\it (ii)} it is {\bf solvable in the strong coupling limit} in the sense that all its correlators can be computed, 
and {\it (iii)} it is {\bf maximally chaotic} in the sense of saturating the Maldacena-Shenker-Stanford (MSS) bound on the Lyapunov 
exponent \cite{Maldacena:2015waa}. While any one of these would certainly justify the attention it has received, 
this confluence of properties makes it an ideal laboratory to explore issues such as the quantum to classical
transition for chaotic systems, information scrambling in black holes, and low-dimensional quantum gravity. 
We will focus on the first of these in this article.\\

\noindent
Toward this end, let us first make some comments about the definition of chaos. Classical chaos was first formally defined on an interval, with the aim of providing a definition that
excludes integrable models, which have an infinite set of commuting conserved charges compatible with the physical degrees of freedom of the system. In this case, there are three
definitions that are usually taken to be equivalent. These are: ({\it i}) a sensitive dependence on initial conditions, ({\it ii}) a transitive mapping function, and ({\it iii}) a dense set
of periodic points. Of these, sensitive dependence on initial conditions implies that any infinitesimally different initial conditions lead to a totally divergent evolution of the system.
Nevertheless, this condition itself does not imply irregular motion with unstable orbits nor a system with rapid memory loss of initial conditions. Therefore, to define classical chaos, we need
transitivity and a dense set of periodic points. Initial studies of classically chaotic dynamical systems treated all three conditions as independent, but it was soon realized that the
sensitive dependence on initial conditions can be derived from, and is in this sense secondary to, the other conditions defining classical chaos on the interval.\\

Of course, since the uncertainty principle negates the concept of conventional trajectories in quantum mechanics, the irregular motion of a quantum particle is also difficult to define.
Nevertheless, the correspondence principle primes us to expect quantum mechanics to have at least something to say about chaos at macroscopic scales. Traditionally, the measure
of choice to diagnose quantum chaos has been the spectral statistics of the quantum system, with a chaotic system exhibiting an irregular spectrum \cite{Percival:1973}. In the
semi-classical limit, this results in a spectrum of discrete bound states with a statistical distribution and repulsion between energy levels. Part of the reason that quantum chaos is
difficult is that, as shown by Berry in his seminal work on the subject \cite{Berry:1977zz}, the classical and integrable limits of a chaotic system do not commute at the
semi-classical level. This semi-classical study also showed that the eigenfunctions of a Hamiltonian of a non-integrable system should display the statistical properties of Gaussian
random coefficients \cite{Berry:1977}, in contrast to an integrable system whose wavefunction cannot be a Gaussian random function. Hence spectral statistics distinguishes between
integrable and non-integrable systems in a non-random Hamiltonian.\\

More recently, several other chaos diagnostics have been proposed, primarily in the context of the condensed matter physics of the SYK model and its variants. Of these, two stand
out. The {\it out-of-time-order correlator} (OTOC),
$C_{T}(t)\equiv \langle[A(t),B(0)]^{2}\rangle_{T}$, with the Heisenberg operators $A(t)$ and $B(t)$ and 
thermal average $\langle\ldots\rangle_{T}$. Originally introduced in the context of 
superconductivity in 1969 \cite{Larkin:1969},  the OTOC has recently been repurposed in the context of 
chaotic many-body systems where $C_{T}(t)\sim \hbar^{2}e^{2\lambda t}$ 
with $\lambda$ identified as the {\it quantum Lyapunov exponent}. For the purposes of our study, it will be 
more useful to focus on another diagnostic; the {\it spectral form factor} recently studied in the Gaussian random 
four-fermion model \cite{Cotler:2016fpe} and random matrix theory \cite{Okuyama:2018yep, Guhr:1997ve} where it was shown that it displays a characteristic dip-ramp-plateau behavior. \\

The central question we would like to address in this article is: {\it How do we distinguish between randomness and chaos?} Having set out how chaos affects the spectrum of a
quantum system, we note that introducing {\it randomness} into the Hamiltonian of an otherwise integrable model has a similar effect on the eigenfunctions. Because the system is
integrable, it should be possible to distinguish, in a controlled way, the effect of randomness from chaos. In what follows, we will use the simplest variant of the SYK system, the
so-called SYK$_{2}$ model, to argue that it is the random eigenvectors of the system that is responsible for the dip-ramp-plateau behavior of the spectral form factor, not any
notion of non-integrability. The system is simple enough that it gives a clear understanding of the physics of randomness and chaos, unobscured by the complexities of the $q>2$
SYK models, where $q$ is the number of interacting Majorana fermion terms.

\section{Spectral Form Factors}
\label{sec:2}
As argued in \cite{Dyer:2016pou}, the spectral form factor 
\bea
g(t)=  \frac{\left|Z(\beta,t)\right|^2}{\left|Z(\beta,0)\right|^2} \,, 
\eea
furnishes a useful diagnostic tool in understanding the chaotic behavior of a many-body system. Here $Z(\beta,t)\equiv\Tr \bigg( \exp\big(-(\beta-it)H\big) \bigg)$ is the un-normalized thermal average of the operator $\exp(it H)$. The variables $\beta$, $t$, and $H$ are the inverse temperature, time and Hamiltonian of the system respectively. In particular, a chaotic system is expected to display a characteristic dip-ramp-plateau behavior in the spectral form factor \cite{Cotler:2016fpe}. In what follows, we compute the spectral form factor in the SYK$_2$ model of two Majorana fermions interacting through a Gaussian random coupling constant. Among its many properties is the fact that it is integrable, leading to two different definitions of parameter-averaged spectral form factor \cite{Cotler:2016fpe}. These are the {\it annealed} spectral form factor. Note that, for the purposes of comparison to the quenched spectral form factor, the annealed spectral form factor is defined from the normalization at the initial time $t=0$. A different definition stated in (\ref{eqn:ann_spec}) is used instead of $g_{\mathrm{ann}}(t)\equiv\langle|Z(\beta, t)|^2\rangle_R/\langle |Z(\beta, 0)|\rangle^2_R$ defined in \cite{Cotler:2016fpe}. The annealed spectral form factor is
\be
g_{\mathrm{ann}}(t)\equiv\frac{\left\langle|Z(\beta, t)|^2\right\rangle_R}{\left\langle|Z(\beta, 0)|^2\right\rangle_R}
\label{eqn:ann_spec}
\ee
and its quenched counterpart is
\be
g_{\mathrm{que}}(t)\equiv\left\langle \left|\frac{Z(\beta, t)}{Z(\beta,0)}\right|^2 \right\rangle_R \,,
\ee
where $\langle {\cal O}\rangle_{R}$ denotes the observable ${\cal O}$ averaged over the random parameters in a system of interest.

\section{SYK$_2$ Model}
\label{sec:3}
The Hamiltonian of the SYK$_2$ model is given by 
\bea
H_{\mathrm{SYK2}}\equiv \frac{i}{2}\sum_{j_1, j_2=1}^N{\cal J}_{j_1 j_2}\psi_{j_1}\psi_{j_2},
\eea
where the real couplings ${\cal J}_{j_1j_2}=-{\cal J}_{j_2j_1}$ are drawn from the Gaussian distribution,
\bea
 \mathcal{P} = \exp\left(-\sum_{j_1,j_2=1}^N{\cal J}_{j_1j_2}^2\frac{N}{4J^2}\right) \,,
\eea
 with zero mean and variance $J^2/ N$. Here $N$ is the number of Majorana fermion fields $\psi$, and $J$ is the coupling constant which is set to unity in this paper. The Majorana fermions satisfy the anti-commutation relation $\{\psi_j, \psi_k\}=\delta_{jk}$.
\\

\begin{figure}[h]
\begin{centering}
\includegraphics[scale=1.]{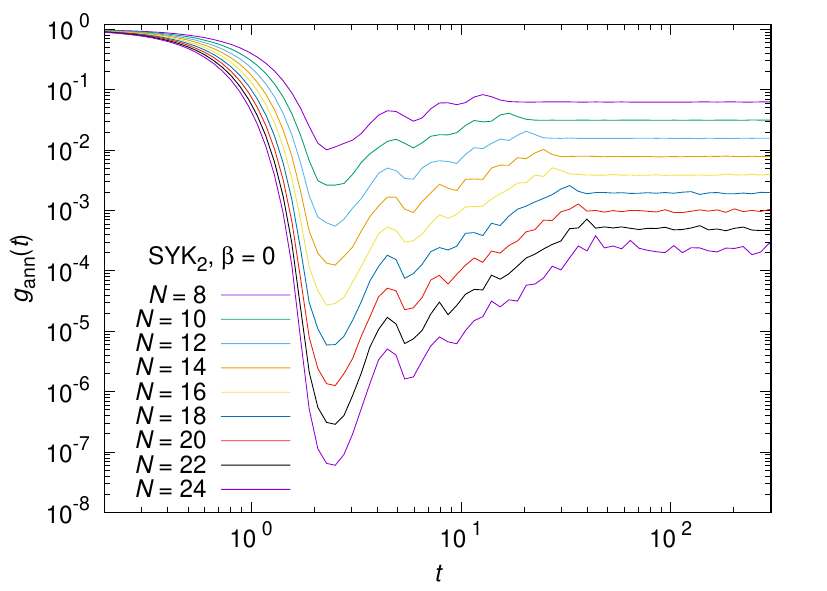}
\par\end{centering}
\caption{This plot shows the annealed spectral form factor $g_{\mathrm{ann}}(t)$ at $\beta=0$ for different numbers of the fermions ranging from $N=8$ to $N=24$. This displays the dip-ramp-plateau behavior in this model. We fix the number of eigenvalues $2^{20}\times 3^3$ to numerically calculate the spectral form factor for each parameter. The minimum number of the Gaussian random configurations is $2^{8}\times 3^3$ for $N=24$.} %The axes are written as the log scale, where $\log 10=1$.}
\label{SYK2-N8-24-SFF-1-anneal-b0.pdf}
\end{figure}

\begin{figure}[h]
\begin{centering}
\includegraphics[width=0.45\textwidth]{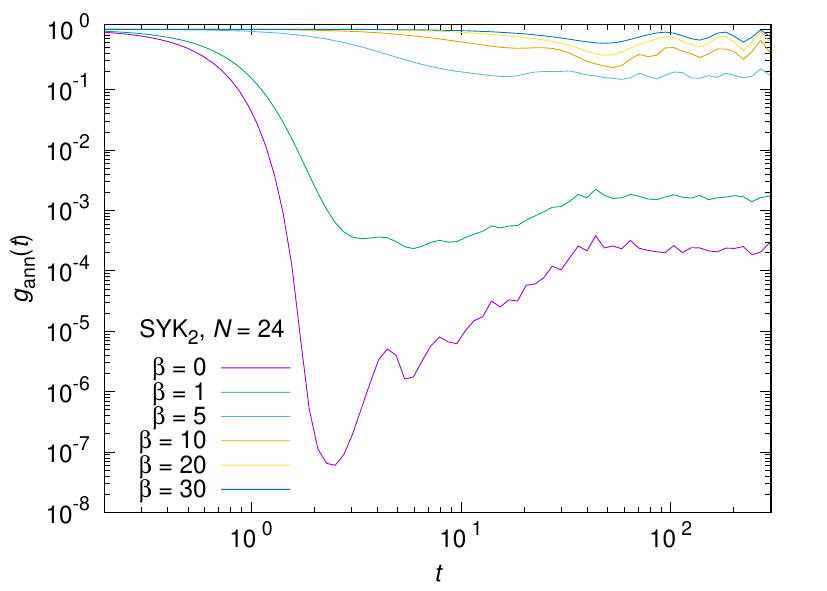}
%\hfill
\includegraphics[width=0.45\textwidth]{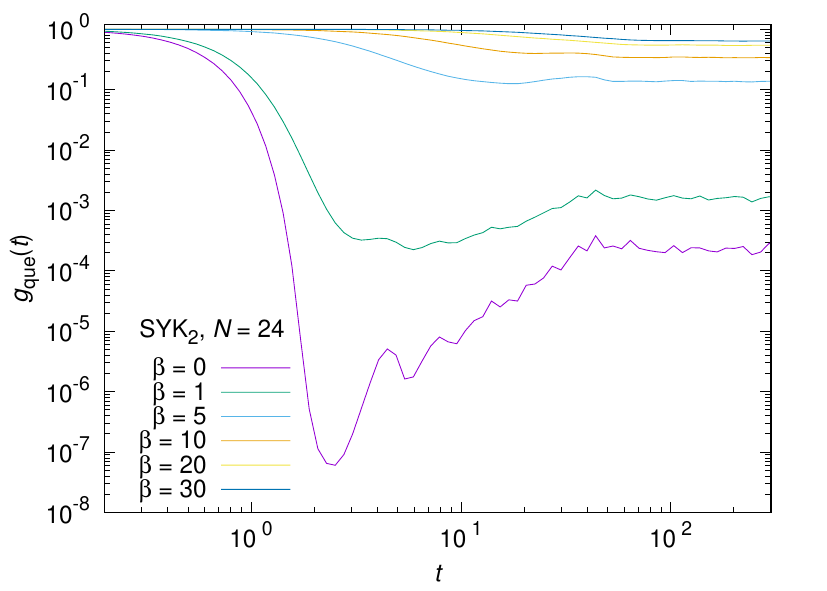}
\par\end{centering}
\caption{We fix $N=24$ and compute the annealed spectral form factor $g_{\mathrm{ann}}(t)$ and the quenched spectral form factor $g_{\mathrm{que}}(t)$ for a range of the inverse temperature from $\beta=0$ to $\beta=30$. Both spectral form factors demonstrate almost identical behavior and display the dip-ramp-plateau behavior at high temperature. A total of $2^{8}\times 3^3$ Gaussian random configurations is used. %The axes are written as the log scale, where $\log 10=1$.
}
\label{SYK2-N8-24-SFF-2-anneal-normalized-N24.pdf}
\end{figure}

We are interested in understanding the relationship between the behavior of the spectral form factor, the number of the fermions, and the inverse temperature $\beta$ of the SYK$_2$ model.
To this end, we compute and plot the annealed spectral form factors at $\beta=0$ in Fig. \ref{SYK2-N8-24-SFF-1-anneal-b0.pdf} for different numbers of fermions.
The annealed spectral form factor $g_{\mathrm{ann}}(t)$ coincides with the quenched spectral form factor $g_{\mathrm{que}}(t)$ at $\beta=0$.
This is to be expected because the partition function at the initial time $t=0$ does not depend on the random coupling constants.
Although this model is integrable, we note that the spectral form factors still display the celebrated dip-ramp-plateau behavior due to the random couplings  
       ${\cal J}$. Note also that the dip becomes more prominent as $N$ or the temperature ($1/\beta$) increases. 
        Fig. \ref{SYK2-N8-24-SFF-2-anneal-normalized-N24.pdf} exhibits the temperature dependence of the spectral form factors for a fixed value of $N=24$.
        There is no significant difference between the annealed and quenched spectral form factors from the numerical study of $N=24$ for some values of the inverse 
        temperature.
\\

\begin{figure}[h]
\begin{centering}
\includegraphics[width=0.45\textwidth]{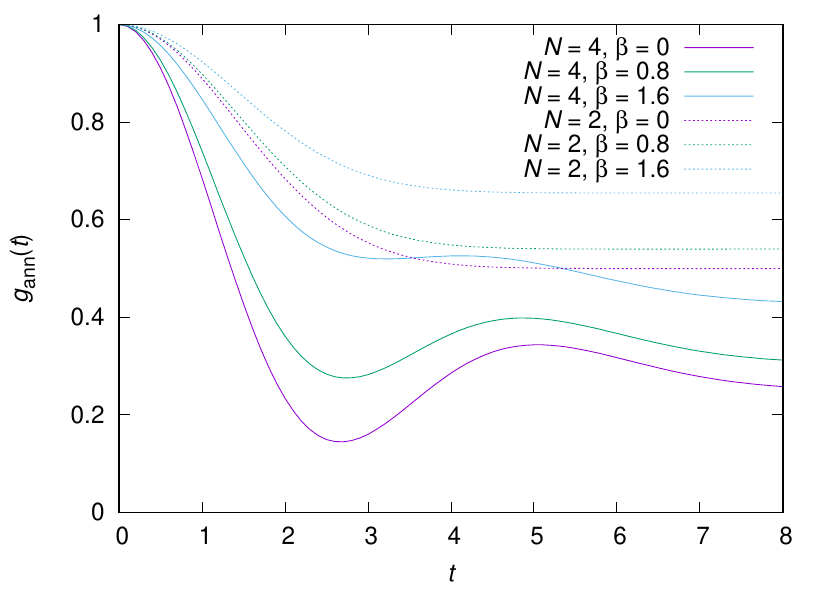}
\hfill
\includegraphics[width=0.45\textwidth]{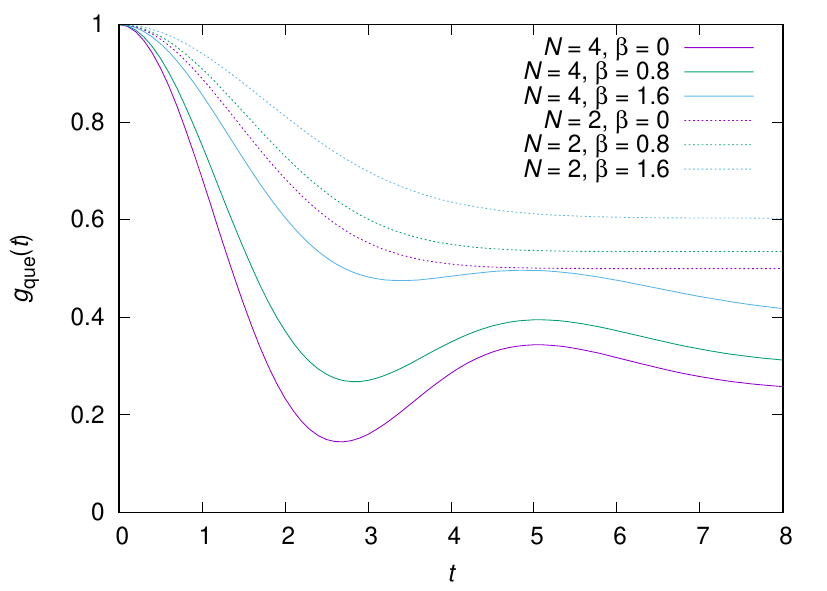}
\par\end{centering}
\caption{We plot both the annealed and quenched spectral form factors for $N=2$ and $N=4$ with $\beta$ ranging from $0$ to $1.6$. Both spectral form factors show the dip-ramp-plateau behavior in the $N=4$ model but not in $N=2$. We fix the number of eigenvalues $2^{20}\times 3^3$ to numerically calculate the spectral form factors for each parameter, The minimum number of the Gaussian random configurations is $2^{18}\times 3^3$ for $N=4$.
}
\label{SYK224SFF.pdf}
\end{figure}

Our numerical analysis of the spectral form factors of the SYK$_2$ model was constrained to values of $N$ with $8 \leq N \leq 24$.
Smaller values of $N$ for the SYK$_2$ model also display a dip-ramp-plateau behavior which persists down to $N=4$.
Surprisingly, the simplest model with $N=2$, as displayed in Fig. \ref{SYK224SFF.pdf} is special and does not display the same dip-ramp-plateau behavior
 To understand why, it is instructive to rewrite the $N=2$ model as a one-qubit system by choosing $\psi_1=\sigma_x/\sqrt{2}$ and $\psi_2=\sigma_y/\sqrt{2}$,
 where as usual, $\sigma_x$ and $\sigma_y$ are Pauli matrices \cite{PM}.
 The corresponding one-qubit Hamiltonian is then simply proportional to $\sigma_z$.
 Given the simple form of the Hamiltonian, one can construct non-trivial conserved charges $Q \propto \sigma_z$. This is no longer possible at larger $N$.
 Another observation is that only the $N=2$ model has constant eigenvectors which are independent of the random parameters. This is also not true in general for larger values of $N$. 
From the perspective of a model with fixed parameters, the random averaging procedure allows us to explore a large set of eigenstates in the model and hence generates the
dip-ramp-plateau behavior in the spectral form factor.\\

The SYK$_2$ model is random but exhibits no repulsion in its spectrum. Consequently, the spectral form factor is not expected to be the same as in random matrix theory. Hence the most precise statement that we can make is that our result only suggests that randomness can generate a dip-ramp-plateau behavior in the spectral form factors without necessarily diagnosing the random matrix theory behavior.  Since integrability can be defined at any $N$, our observations for small $N$ motivates an interesting diagnosis of integrability without restricting to large $N$ properties of the system. To obtain more evidence in support of this conjecture, we next study a Gaussian random transverse Ising model. 

\section{Gaussian Random Transverse Ising Model}
\label{sec:4}
The Hamiltonian of the Gaussian random transverse Ising model is
\bea
H_{\mathrm{ti}}=-\!\!\!\!\!\!\!\!\!\!\!\!\sum_{i_1=1,2,\cdots, L-1}\!\!\!\!\!\!\!\!K_{i_1}\sigma^z_{i_1}\sigma^z_{i_1+1}-M\!\!\!\!\!\!\!\!\sum_{i_2=1,2,\cdots ,L}\!\!\!\!\!\!\!\!\sigma^x_{i_2},
%\nn\\
\eea
in which the random coupling constants follow the Gaussian distribution with zero mean and unit variance. The number of the lattice sites is $L$. The superscript of Pauli matrix denotes the component of Pauli matrix and the subscript of Pauli matrix denotes the lattice site.\\ 

\begin{figure}
\begin{centering}
	\begin{subfigure}{1.\textwidth}
		\includegraphics[width=0.45\textwidth]{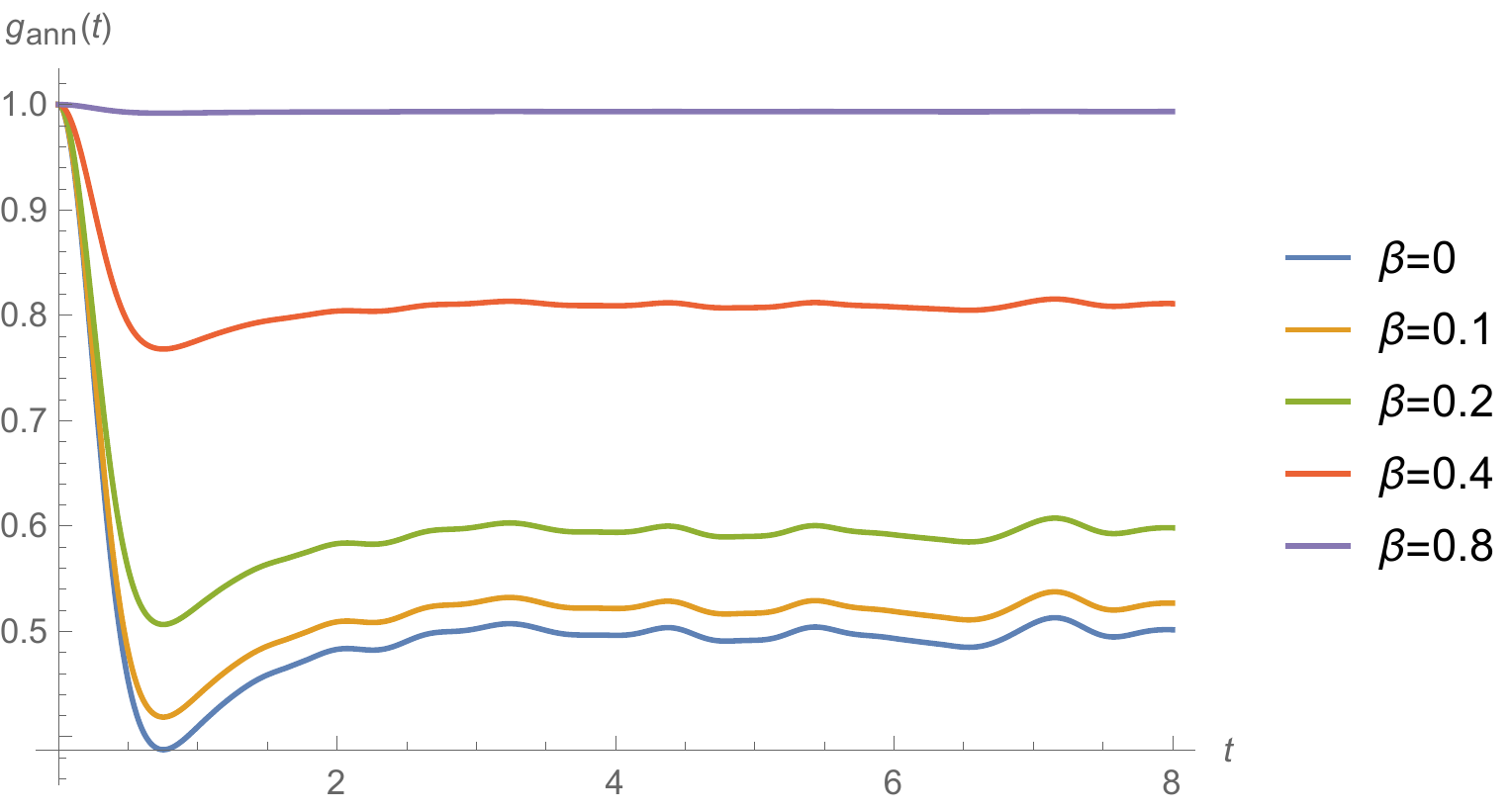}
		\hspace{0.1cm}
		\includegraphics[width=0.45\textwidth]{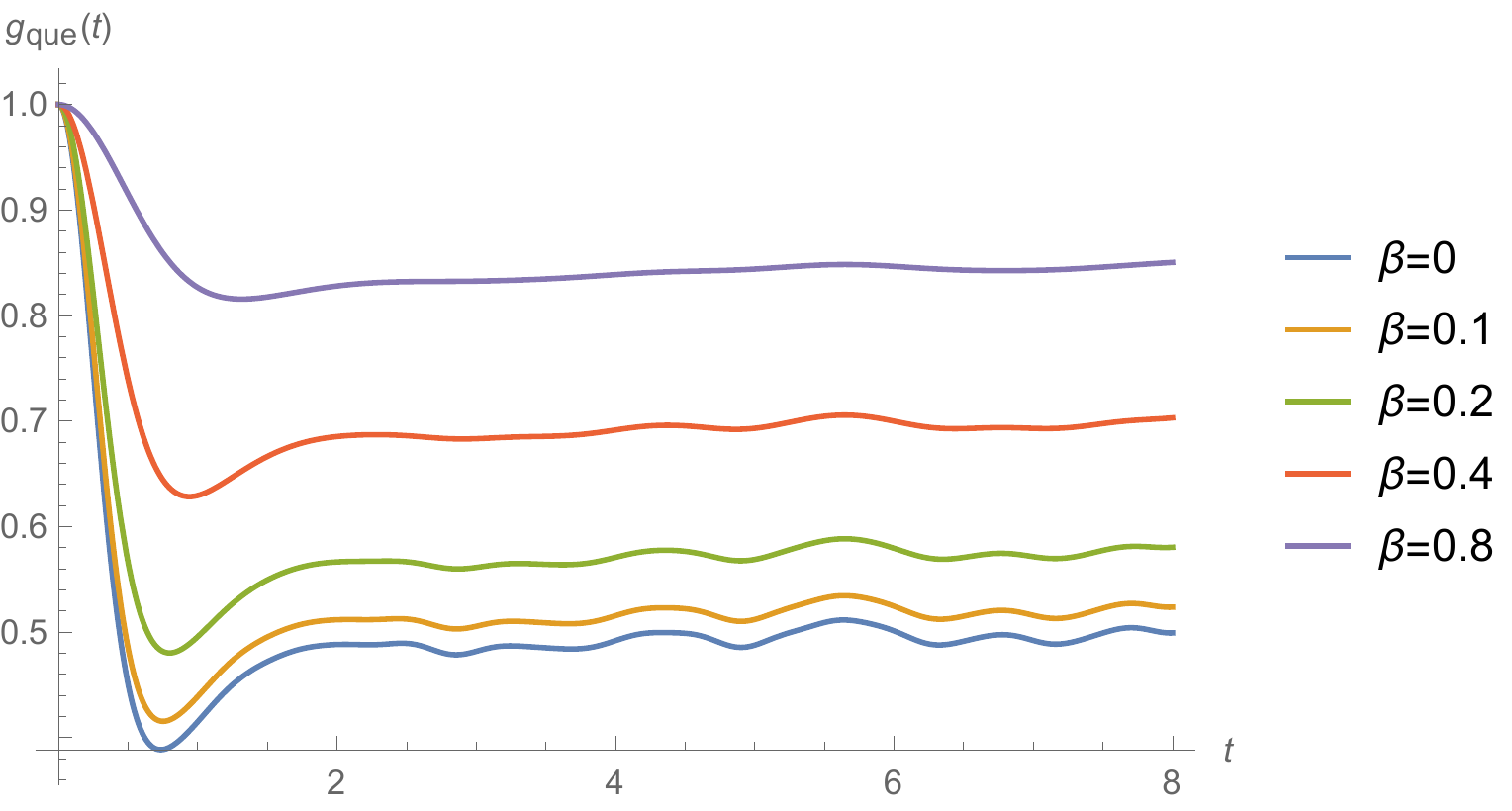}
	\caption{$L$=2}
	\end{subfigure}\\
	\begin{subfigure}{1.\textwidth}
		\includegraphics[width=0.45\textwidth]{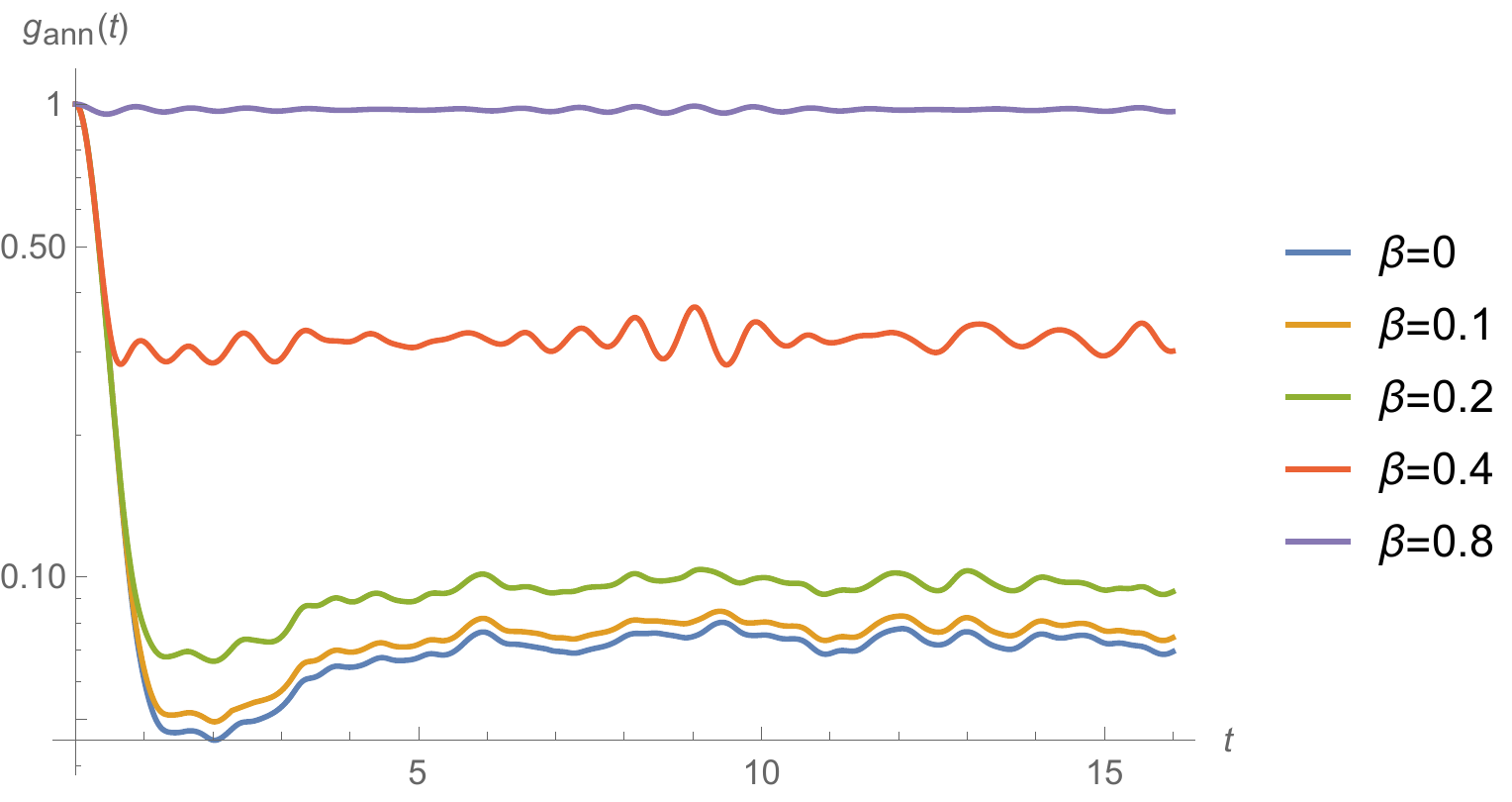}
		\hspace{0.1cm}
		\includegraphics[width=0.45\textwidth]{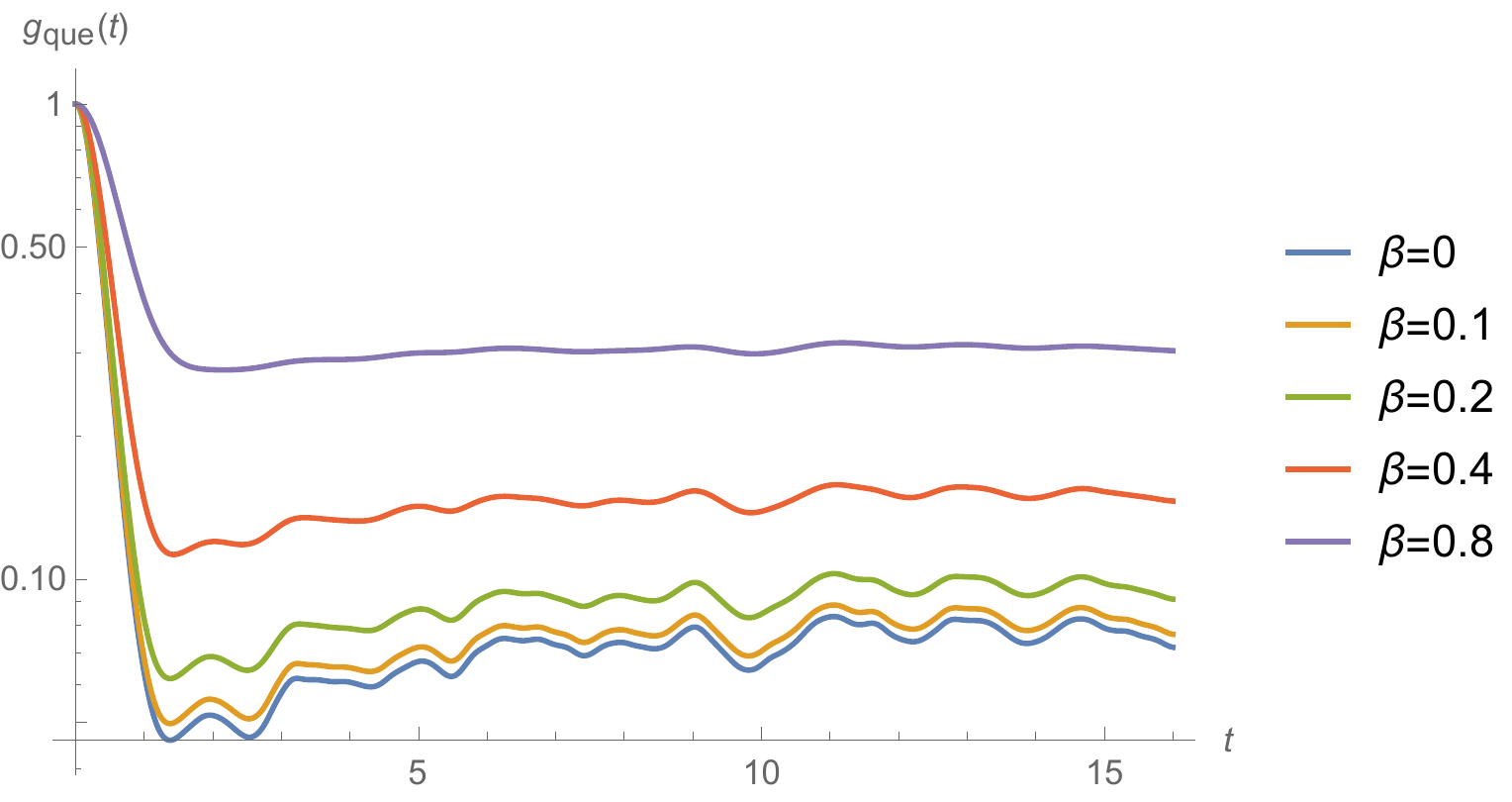}
	\caption{$L$=4}
	\end{subfigure}
\par\end{centering}
	\caption{We compute the annealed spectral form factor $g_{\mathrm{ann}}(t)$ and the quenched spectral form factor $g_{\mathrm{que}}(t)$ for a range of the inverse temperatures from $\beta=0.1$ to $\beta=0.8$ for the case of $L=2$ and $L=4$. Both spectral form factors display the dip-ramp-plateau behavior at high temperature. A total of $2048$ Gaussian random configurations is used. 
	}
	\label{ti2.pdf}
\end{figure}

We show the result in the annealed and quenched spectral form factors in Fig. \ref{ti2.pdf} for $L=2$ and $L=4$ at the finite temperatures $\beta$=0, 0.1, 0.2, 0.4, and 0.8.  These figures show the dip-ramp-plateau behavior at a high temperature regime from the 2048 Gaussian random configurations without suffering from the finite $N$ artifact.
%\begin{comment}
%\begin{figure}[h]
%\begin{centering}
%\includegraphics[width=0.23\textwidth]{ti2specm.pdf}
%\hspace{0.1cm}
%\includegraphics[width=0.23\textwidth]{ti2specmm.pdf}
%\par\end{centering}
%\caption{We fix $L=2$ and compute the annealed spectral form factor $g_{\mathrm{ann}}(t)$ and the quenched spectral form factor $g_{\mathrm{que}}(t)$ for a range of the inverse temperatures from $\beta=0.1$ to $\beta=0.8$. Both spectral form factors display the dip-ramp-plateau behavior at high temperature. A total of $2048$ Gaussian random configurations is used. 
%}
%\label{ti2.pdf}
%\end{figure}
%\end{comment}
%\begin{comment}
%\begin{figure}[h]
%\begin{centering}
%\includegraphics[width=0.23\textwidth]{ti4specm.pdf}
%\hspace{0.1cm}
%\includegraphics[width=0.23\textwidth]{ti4specmm.pdf}
%\par\end{centering}
%\caption{We fix $L=4$ and compute the annealed spectral form factor $g_{\mathrm{ann}}(t)$ and the quenched spectral form factor $g_{\mathrm{que}}(t)$ for a range of the inverse temperatures from $\beta=0.1$ to $\beta=0.8$. Both spectral form factors display the dip-ramp-plateau behavior at high temperature. A total of $2048$ Gaussian random configurations is used. %The value of the spectral form factors are written as the log scale.
%}
%\label{ti4.pdf}
%\end{figure}
%\end{comment}

\section{Decoherence}
\label{sec:5}
Various studies of chaotic systems suggest that they are closely related to the properties of thermal ensembles \cite{Srednicki:1994}.
This section illustrates this relation in the few-body system explicitly using the Gaussian random one-qubit model.
We show that the expectation values of some observables in a specific random quantum state approach the thermal ensemble values.\\

Let us start with the most general quantum state
\begin{equation}
\left|\psi(t)\right\rangle=\left(\sum_{j=1}^{2^n}|a_j|^2\right)^{-1}\sum_{k=1}^{2^n}a_ke^{-i\lambda_k t}\left|\psi_k\right\rangle,
\label{eqn:gen_stat}
\end{equation}
where $\left|\psi_k\right\rangle$ is the $k$-th eigenstate and $\lambda_k$ is the corresponding eigenvalue.
The expectation value of a physical observable ${\cal O}$ for this state is given by
\bea
&&\left\langle\psi(t)\right|{\cal O}\left|\psi(t)\right\rangle
%\nn\\
=\frac{\mathrm{Tr} \left(e^{-\beta H_{\mathrm{oq}}}{\cal O}\right) }{\mathrm{Tr}\ e^{-\beta H_{\mathrm{oq}}}}
%\nn\\
+\Bigg(\sum_{j=1}^{2^n}|a_j|^2\Bigg)^{-1}
\nn\\
&&\times
\left(\sum_{l_1, l_2=1,l_1\ne l_2}^{2^n}a_{l_1}^*a_{l_2}e^{i(\lambda_{l_1}-\lambda_{l_2})t}\left\langle\psi_{l_1}\right|{\cal O}\left|\psi_{l_2}\right\rangle\right) \,,
\label{eqn:state_avg}
\eea
where the first term in the second line is the thermal expectation value of $\cal O$. This relation can be established if one chooses $|a_k|^2=\exp \left( -\beta\lambda_k \right) $. We demonstrate the relation using the Gaussian random one-qubit model below.\\

The Hamiltonian of the Gaussian random one-qubit model is $H_{\mathrm{oq}}\equiv g_x\sigma_x+g_y\sigma_y+g_z\sigma_z \,,$
where the parameters $g_x$, $g_y$, and $g_z$ are drawn from a Gaussian distribution with zero mean and 
unit variance. This model is integrable and has eigenvalues 
$\lambda_{\pm}=\pm\sqrt{g_x^2+g_y^2+g_z^2}$ with corresponding eigenvectors $|\psi_{\pm}\rangle=
\begin{pmatrix}
	1 & a_{\pm} 
\end{pmatrix}^T\big/\sqrt{1+|a_{\pm}|^2} \,,$
where $a_{\pm}\equiv (g_x-ig_y)(\lambda_{\pm}+g_z)/(g_x^2+g_y^2)$, and $T$ is the transpose 
operation. In the Gaussian random one-qubit model, there are three non-trivial observables $\sigma_x$, $\sigma_y$, and $\sigma_z$. We are interested in the time evolution of the expectation value of these observable. We focus on the second term in \eqref{eqn:state_avg} and label them as ($A_1(t)$,$A_2(t)$,$A_3(t)$) for ${\cal O}=(\sigma_x,\sigma_y,\sigma_z)$, respectively. With the specific choice of $a_k=\exp({-\beta\lambda_k/2})$, one obtains
\begin{align}
A_1(t)&= \bigg\langle\frac{1}{e^{-\beta\lambda_+}+e^{\beta\lambda_+}}
\nn\\
&\times\bigg(\frac{2g_xg_z}{\lambda_+\sqrt{\lambda_+^2-g_z^2}}\cos(2\lambda_+t)-\frac{2g_y}{\sqrt{\lambda_+^2-g_z^2}}\sin(2\lambda_+t)\bigg)\bigg\rangle_{R},
\nn\\
A_2(t)&=\bigg\langle\frac{1}{e^{-\beta\lambda_+}+e^{\beta\lambda_+}}
\nn\\
&\times\bigg(\frac{2g_yg_z}{\lambda_+\sqrt{\lambda_+^2-g_z^2}}\cos(2\lambda_+t)+\frac{2g_x}{\sqrt{\lambda_+^2-g_z^2}}\sin(2\lambda_+t)\bigg)\bigg\rangle_{R},
\nn\\
A_3(t)& =\bigg\langle\frac{1}{e^{-\beta\lambda_+}+e^{\beta\lambda_+}}\frac{2\sqrt{g_x^2+g_y^2}}{\lambda_+}\cos(2\lambda_+t)\bigg\rangle_{R}.
\end{align}
\\

\begin{figure}
\begin{centering}
\includegraphics[scale=0.3]{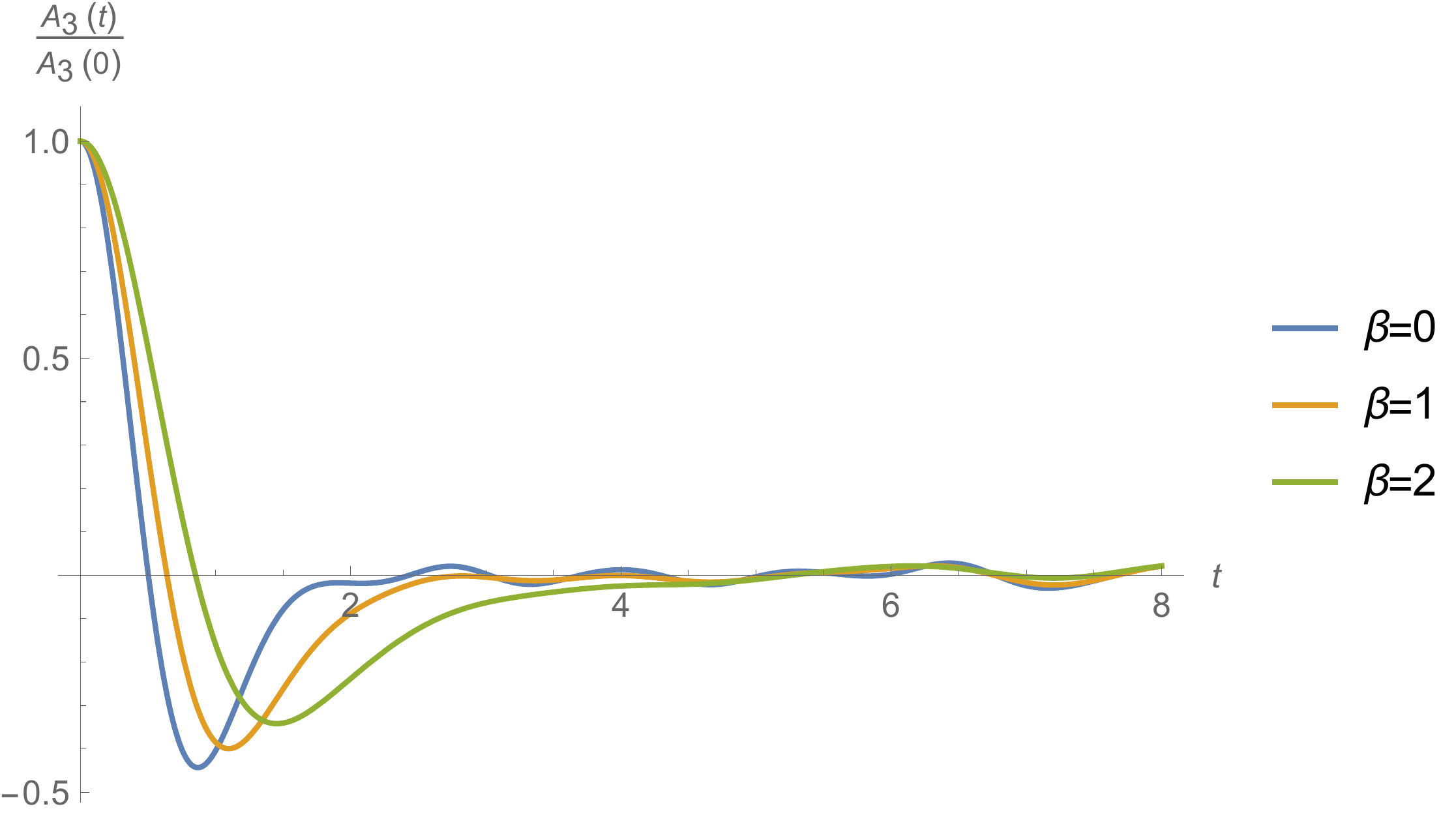}
\par\end{centering}
\caption{We plot the normalised function $A_3(t)/A_3(0)$ at $\beta=0, 1,2$. The decoherence phenomenon happens earlier at higher temperature. We use $2^{11}$ Gaussian random configurations for this numerical computation.
}
\label{dec1.pdf}
\end{figure}

The quantities $A_1(t)$ and $A_2(t)$ vanish for all $t$ and $A_3(t)$ decays to $0$ in this model (See Fig. \ref{dec1.pdf}). This result shows that the density matrix of \eqref{eqn:gen_stat} approaches the thermal ensemble upon averaging over the random parameters. The decoherence rate is faster at a higher temperature.\\

Note that our result does not produce eigenstate thermalization \cite{Srednicki:1999} since the latter also requires exponential decay with increasing system size. Randomness is however enough to provide a thermal ensemble.

\section{Outlook}
\label{sec:6} 
Through a detailed analytic and numerical study, we have provided extensive evidence to support our claim that the dip-ramp-plateau behavior of the spectral form factor \cite{Cotler:2016fpe, Dyer:2016pou} is due to randomness instead of chaos in the SYK$_2$ model. Similar behavior was also observed in the Gaussian random transverse Ising model. This sensitivity of the spectral form factor to the randomness of the system suggests that the spectral form factor
can be used as an order parameter of the randomness. We also found that the Gaussian random one-qubit system exhibits a thermal ensemble at late times just as in the case of a
quantum chaotic system. As the randomness provides the statistical distribution of the system, the dip-ramp-plateau behavior implies amnesia of the initial conditions. Hence our
result suggests that the behavior of spectral form factor should be taken as a supplementary condition for quantum chaos, similar to the transitivity of the mapping function in the
case of classical chaos. Therefore, we expect that the dip-ramp-plateau behavior in the spectral form factor should be related to the classical-quantum transition.\\

Our result just suggested that the dip-ramp-plateau behavior in the annealed and quenched spectral form factors and the thermal ensemble can only come from the randomness
without restricting to the integrability. It is still possible to use other quantities to study chaos. For example, the connected unfolded spectral form factor \cite{Nosaka:2018iat}.
This quantity does not provide the dip-ramp-plateau behavior to the SYK$_2$ model and the Gaussian random one-qubit model, but the Gaussian random one-qubit model has the
Gaussian unitary ensemble \cite{Dyson:1962es}.
Hence the precise determination of the chaos needs a more careful study when one includes the random Hamiltonian into the study.\\

The SYK$_2$ model is a non-interacting theory of random fermions. As such, one can diagonalize the single-body Hamiltonian to obtain its eigenvalues and study the large $N$ physics \cite{Dumitriu:2009}. Indeed, the model is similar enough to the random matrix theory that we have hope that the spectral form factor possibly has an exact analytic expression \cite{Okuyama:2018yep}. That said since the random numbers used in computer simulations are not exactly random, and larger $N$ requires larger Gaussian random ensembles to realize these results, a combination of analytical and numerical solutions is both necessary and interesting.\\

 The topic of quantum chaos is enjoying a resurgence in light of recent developments in the SYK model and low-dimensional quantum gravity. As such, much of what was known during its initial development in the 1980s and 1990s has come under the magnifying glass, theorems sharpened and new tools developed in the context of new analytic models and computational advances like the conformal bootstrap. Our analysis of the Gaussian random integrable system that is the SYK$_2$ model suggests that even in simple systems like this, the exploration of the various ingredients that constitute quantum chaos is still in its infancy, with exciting times ahead.

\section*{Acknowledgments}
We would like to thank Paolo Glorioso, Aitor Lewkowycz, Wolfgang Mück, Laimei Nie, Xiao-Liang Qi, Dario Rosa, Shinsei Ryu, and Stephen H. Shenker for useful discussions.
PHCL was supported by the Croucher Fellowship. CTM was supported by the Post-Doctoral International Exchange Program. JM is supported in part by the NRF of South Africa under grant CSUR 114599 and the National Science Foundation under Grant No. NSF PHY-1748958. MT was partially supported by a Grant-in-Aid No. JP17K17822 from JSPS of Japan. CTM would like to thank Nan-Peng Ma for his encouragement. We would like to thank the National Tsing Hua University, Tohoku University, Okinawa Institute of Science and Technology Graduate University, Yukawa Institute for Theoretical Physics, Istituto Nazionale Di Fisica Nucleare - Sezione di Napoli, Kadanoff Center for Theoretical Physics, Stanford Institute for Theoretical Physics, the KITP, and Israel Institute for Advanced Studies, for hospitality at various stages of this work.

%\appendix

\end{document}